\documentclass[prb,twocolumn,showpacs,preprintnumbers,amsmath,amssymb]{revtex4}

\usepackage{graphicx}
\usepackage{dcolumn}
\usepackage{bm}

\renewcommand{\vec}{\mathbf} 
\newcommand{\lgfk} {\langle \! \langle } 
\newcommand{\rgfk} {\rangle \! \rangle } 

\begin{document}

\preprint{}

\title{Electronic properties of EuB$_{6}$ in the ferromagnetic regime:
    Half-metal versus semiconductor}

\author{M. Kreissl}

\email{kreissl@physik.hu-berlin.de}

\author{W. Nolting}

\affiliation{Lehrstuhl Festk\"{o}rpertheorie, Institut f\"{u}r Physik, Humboldt-Universit\"{a}t
zu Berlin, Newtonstr. 15, 12489 Berlin}

\begin{abstract}
To understand the halfmetallic ferromagnet EuB$_{6}$ we use the Kondo
lattice model for valence and conduction band. By means of a recently
developed many-body theory we calculate the electronic properties
in the ferromagnetic regime up to the Curie temperature. The decreasing
magnetic order induces a transition from halfmetallic to semiconducting
behavior along with a band broadening. We show the temperature dependence
of the quasiparticle density of states and the quasiparticle dispersion
as well as the effective mass, the number of carriers and the plasma
frequency which are in good agreement with the experimental data. 
\end{abstract}

\pacs{71.10.-w, 75.50.Pp, 71.30.+h, 75.10.-b, 75.30.-m}

\maketitle

\section{Introduction}

In the last decade EuB$_{6}$ with its interesting and still not fully
understood magnetic and electronic properties has attracted considerable
interest. For low temperatures it is a ferromagnetic semimetal becoming
a paramagnetic semiconductor for higher temperatures. It crystallizes
in the cubic CaB$_{6}$ structure (spacegroup 221, Pm3m). The divalent
Europium ions occupy the corners of the cubic unit cell and at the
body centered position sits an octahedron consisting of six Boron
atoms. The lattice constant a=4.19\AA{} does not change with
temperature. \cite{key-suellow98}

EuB$_{6}$ exhibits two consecutive phase transitions at T$_{C}$=12.5K
and T$_{M}$=15.5K. \cite{key-suellow98} T$_{C}$ is the bulk Curie
temperature. The magnetization follows a Brillouin function and the
saturation value is 7$\mu_{B}$ due to the exactly half-filled 4f
shell in Europium.\cite{key-magnetization} Nonetheless there are
evidences of phase separation above T$_{C}$ up to 30K \cite{key-suellow2000,key-magnetization},
where bound magnetic polarons \cite{key-nyhus} are responsible for
the cusplike peak in resistivity and the colossal magnetoresistance
at T$_{M}$. \cite{key-urbano,key-brooks} Concomitant with the bulk
magnetic phase transition at T$_{C}$ is a transition from semimetal
to semiconductor. \cite{key-tunnel,key-nyhus}

For low temperatures the semimetallic behavior is proven theoretically
and experimentally. Early local density approximation (LDA) \cite{key-hexaborids}
as well as recent band structure calculations in LDA+U \cite{key-kunes}
show an overlap of valence and conduction band around the X-point
of the Brillouin zone. Shubnikov - de Haas and de Haas - van Alphen
measurements \cite{key-aronson,key-goodrich} agree with that picture
as they report electron and hole pockets on the Fermi surface at the
X-point. The carrier density is about 0.01 per unit cell for low and
0.001 for higher temperatures. \cite{key-paschen} Due to the known
lack of correct predictions of band gaps in the LDA the authors in
{[}\onlinecite{key-kunes}{]} suggest EuB$_{6}$ to be halfmetallic
in the ferromagnetically ordered phase. That scenario is not ruled
out by experiments albeit not proven also.

The resistivity shows a strong peak at T$_{M}$ and also indicates
a metal - insulator transition. Up to T$_{C}$ and T$_{M}$ an increase
of the carrier concentration, a reduction of the effective mass or a 
combination of both is responsible for a strong red shift of the plasma 
frequency.\cite{key-degiorgi97} The specific
electronic properties seem to stem from temperature dependent changes
of valence and conduction bands induced by the ferromagnetic alignment
of the local moments. The other way around the ferromagnetism is attributed
to an RKKY mechanism where the itinerant carriers are responsible
for the magnetic exchange. It looks as if the system will be best
described by the two band Kondo lattice model introduced in the next
section.

Similar models but with focus on other properties of EuB$_{6}$ are
already studied in a mean-field treatment \cite{key-korenblit} and
in dynamical mean-field theory. \cite{key-lin millis} The spin polaron
model \cite{key-chaterjee} explains the behavior in resistivity and
the colossal magnetoresistance effect around T$_{M}$. This paper
will focus on the range from zero temperature to T$_{C}$ where the
magnetic and metallic phase transitions take place.

\section{Theory}

For modeling EuB$_{6}$ we will use the Kondo-lattice model, \cite{key-KLM}
also known as s-f model. The spins of the localized 4f electrons result
in a local magnetic moment which interacts with the spin of the itinerant
electrons in the valence(v) and conduction(c) band, respectively.
The complete two-band Hamiltonian in second quantization is\begin{eqnarray}
H & = & \sum_{b}^{v,c}\sum_{ij\sigma}(T_{ij}^{b}-\mu\delta_{ij})a_{bi\sigma}^{\dagger}a_{bj\sigma}+\nonumber \\
 &  & -\sum_{b,b^{\prime}}^{v,c}J^{bb^{\prime}}\sum_{j\sigma\sigma\prime}\frac{1}{2}a_{bj\sigma}^{\dagger}\tau_{\sigma\sigma^{\prime}}a_{b^{\prime}j\sigma^{\prime}}\vec{S}_{j}.\label{eq:H2band}\end{eqnarray}
 The first term in (\ref{eq:H2band}) describes the kinetic part of
the electrons in band b with the hopping integrals T$_{ij}^{b}$ between
two lattice sites $\vec{R}_{i}$ and $\vec{R}_{j}$ that are connected
to the wave vector $\vec{k}$ dependent Bloch energies $\varepsilon_{\vec{k}}^{b}$
via Fourier transformation $T_{ij}^{b}=\frac{1}{N}\sum_{\vec{k}}\varepsilon_{\vec{k}}^{b}e^{i\vec{k}\vec{R}_{ij}}$.
The chemical potential is $\mu$ and $a_{bj\sigma}^{\dagger}(a_{bj\sigma})$
the creation (annihilation) operator of an electron with spin $\sigma$
at lattice site $\vec{R}_{j}$ in band b.

The second term in (\ref{eq:H2band}) describes the s-f interaction
between the itinerant band electrons and the local moments $\vec{S}_{j}$.
$\tau_{\sigma\sigma\prime}$ are the Pauli matrices and J$^{bb\prime}$
the coupling constants. In contrast to the kinetic part the interaction
term does not need to be diagonal in band indices. It might be possible
that during the s-f interaction the itinerant electron in band b hops
to the other band b'. That possibility is included in (\ref{eq:H2band}).

In EuB$_{6}$ the valence band is formed from Boron 2p and Europium
4f states and the conduction band from Boron 2p and Europium 5d states
which overlap at the X-point \cite{key-kunes}. The possible interband
transitions are negligible by symmetry reasons.

Therefore we will set J$^{bb\prime}$ diagonal, resulting in the following
Hamiltonian \begin{equation}
H=\sum_{b}^{v,c}(H_{s}^{b}+H_{sf}^{b}).\end{equation}
 H$_{s}^{b}$ is the kinetic term for band b \begin{equation}
H_{s}^{b}=\sum_{ij\sigma}(T_{ij}^{b}-\mu\delta_{ij})a_{bi\sigma}^{\dagger}a_{bj\sigma}.\end{equation}
 H$_{sf}^{b}$ describes the s-f interaction for band b \begin{equation}
H_{sf}^{b}=-\frac{1}{2}J^{b}\sum_{j\sigma}(z_{\sigma}S_{j}^{z}n_{bj\sigma}+S_{j}^{\sigma}a_{bj-\sigma}^{\dagger}a_{bj\sigma})\label{eq:Hsf}\end{equation}
 where we used $n_{bj\sigma}=a_{bj\sigma}^{\dagger}a_{bj\sigma}$,
$z_{\sigma}=\delta_{\sigma\uparrow}-\delta_{\sigma\downarrow}$ and
$S_{j}^{\sigma}=S_{j}^{x}+iz_{\sigma}S_{j}^{y}$. The first term in
(\ref{eq:Hsf}) describes an Ising-like interaction and the second
the spin-flip processes.

The coupling constant J$^{b}$ is positive (ferromagnetic alignment)
for the conduction band electrons and negative (anti-ferromagnetic
alignment) for the valence band electrons\cite{key-kunes}. We assume
the same magnitude J for both bands, hence J$^{v}$=-J and J$^{c}$=J.
 
It goes without saying, that the k-independent coupling constant is 
an oversimplification. However, this is unavoidable due to mathematical 
difficulties.

\subsection{Green Function}

We are mainly interested in the single electron Green function\begin{eqnarray}
G_{\vec{k}\sigma}^{b}(E) & = & \lgfk a_{b\vec{k}\sigma};a_{b\vec{k}\sigma}^{\dagger}\rgfk\\
 & = & \frac{1}{N}\sum_{ij}\lgfk a_{bi\sigma};a_{bj\sigma}^{\dagger}\rgfk e^{i\vec{k}\vec{R}_{ij}}\end{eqnarray}
 from which all electronic properties can be calculated. The operator
$a_{b\vec{k}\sigma}^{\dagger}(a_{b\vec{k}\sigma})$ is the creation
(annihilation) operator in k-space. To derive the Green function we
solve the equation of motion\begin{equation}
E\, G_{\vec{k}\sigma}^{b}(E)=\hbar+\lgfk[a_{b\vec{k}\sigma},H_{s}^{b}+H_{sf}^{b}]_{-};a_{b\vec{k}\sigma}^{\dagger}\rgfk.\end{equation}
 For some special cases the equation of motion can be solved exactly
but for most of the cases one needs to apply specific approximations
for the higher Green functions to decouple. Decoupling means it is
possible to introduce a self energy $\Sigma_{\vec{k}\sigma}^{b}(E)$
which formally solves \begin{equation}
\lgfk[a_{b\vec{k}\sigma},H_{sf}^{b}]_{-};a_{b\vec{k}\sigma}^{\dagger}\rgfk_{E}=\Sigma_{\vec{k}\sigma}^{b}(E)\, G_{\vec{k}\sigma}^{b}(E)\end{equation}
 and thus the Green function reads\begin{equation}
G_{\vec{k}\sigma}^{b}(E)=\frac{\hbar}{E-\varepsilon_{\vec{k}}^{b}+\mu-\Sigma_{\vec{k}\sigma}^{b}(E)}.\label{eq:greenfunction}\end{equation}

\subsection{Self Energy}

In this section we will introduce an appropriate approximation of
the self energy for the almost empty conduction band, which is ferromagnetically
coupled to the local magnetic moments, and the almost filled valence
band which is anti-ferromagnetically coupled.

For the \textbf{conduction band} it is possible to apply the interpolation
formula for the self-energy given in {[}\onlinecite{key-lowdensity}{]}
which is valid for low densities (e$\rightarrow$0) and covers the
following exactly solvable special cases:

\begin{itemize}
\item Zero-bandwidth limit ($\varepsilon_{\vec{k}}\rightarrow T_{0})$
\item Ferromagnetically saturated semiconductor\\ ($\langle S^{z}\rangle$=S,
e=0)
\item Second order perturbation theory in J
\item High energy expansion
\end{itemize}
The resulting self-energy is\begin{eqnarray}
\lefteqn{\Sigma_{\sigma}^{c}=\Sigma_{\sigma}^{e\rightarrow0}(E)=}\nonumber \\
 &  & -\frac{1}{2}J^{c}\Bigl(M_{\sigma}-\frac{1}{2}J^{c}\frac{a_{\sigma}G_{0}(E-\frac{1}{2}J^{c}M_{\sigma})}{1-\frac{1}{2}J^{c}G_{0}(E-\frac{1}{2}J^{c}M_{\sigma})}\Bigr)\label{eq:selfenergycondband}\end{eqnarray}
 where $M_{\sigma}=z_{\sigma}\langle S^{z}\rangle$, $a_{\sigma}=S(S+1)-m_{\sigma}(m_{\sigma}+1)$
and $G_{0}(E)=\frac{1}{N}\sum_{\vec{k}}\frac{\hbar}{E-\varepsilon_{\vec{k}}+\mu}$
is the free propagator. The magnetization $\langle S^{z}\rangle$
will be calculated via the Brillouin function which is in agreement
with the measurements in {[}\onlinecite{key-magnetization}{]}.

Since the \textbf{valence band} is almost completely filled, let us
change for one moment from the electronic picture to a hole like quasiparticle
picture. Such a hole corresponds to a missing electron in an otherwise
completely filled valence band. To keep the negative coupling constant
for the electrons in the valence band, the coupling constant for the
holes must have opposite sign ($J^{v}(e)\rightarrow-J^{v}(h)$). The
self energy for the holes in the low density approximation (h$\rightarrow$0)
from {[}\onlinecite{key-lowdensity}{]} then reads\begin{eqnarray}
\lefteqn{\Sigma_{\sigma}^{h\rightarrow0}(E)=}\nonumber \\
 &  & +\frac{1}{2}J^{v}\Bigl(M_{\sigma}+\frac{1}{2}J^{v}\frac{a_{\sigma}G_{0}(E+\frac{1}{2}J^{v}M_{\sigma})}{1+\frac{1}{2}J^{v}G_{0}(E+\frac{1}{2}J^{v}M_{\sigma})}\Bigr).\end{eqnarray}
 Returning to the electron picture through reversing the spin ($M_{\sigma}(h)\rightarrow-M_{\sigma}(e)$),
the self energy for the electrons in the almost completely filled
valence band then is\begin{eqnarray}
\lefteqn{\Sigma_{\sigma}^{v}(E)=\Sigma_{-\sigma}^{h\rightarrow0}(E)=}\nonumber \\
 &  & -\frac{1}{2}J^{v}\Bigl(M_{\sigma}-\frac{1}{2}J^{v}\frac{a_{-\sigma}G_{0}(E-\frac{1}{2}J^{v}M_{\sigma})}{1+\frac{1}{2}J^{v}G_{0}(E-\frac{1}{2}J^{v}M_{\sigma})}\Bigr).\label{eq:selfenergyvalband}\end{eqnarray}
 In fact this self-energy covers the above mentioned special cases
for full bands.

Now we can put the selfenergies for both bands, (\ref{eq:selfenergycondband})
and (\ref{eq:selfenergyvalband}) in (\ref{eq:greenfunction}) and
calculate the Greenfunction.

\section{Results\label{sec:Results}}

As described in the previous section we are able to calculate the
one electron Green function with the given self energies for both
bands. In this section our results are presented. The band index will
be skipped as it should be clear that the following calculations are
formal equivalent for both bands.

Several free parameters need to be adapted. The spin is S=$\frac{7}{2}$
due to the exactly half filled 4f shell of Europium and the coupling
constant J=0.2eV is in accordance with current literature. We assume
a modified tight binding dispersion $\varepsilon_{\vec{k}}$ for simple
cubic\cite{key-tight-binding} with the band gap located at the X-point
instead of $\Gamma$.\begin{equation}
\varepsilon_{\vec{k}}=T_{0}-\frac{W}{6}\Bigl(cos(k_{x}a-\pi)+cos(k_{y}a)+cos(k_{z}a)\Bigr)\end{equation}
Here T$_{0}$ is the band's center of gravity and W the bandwidth.
The widths of the
tight binding bands are crucial for our model calculation since the
effective coupling strength is $\frac{JS}{W}.$ The mean values of
the anisotropic effective masses m$^{*}$ given in {[}\onlinecite{key-kunes}{]}
are 1.2m$_{e}$ and 0.35m$_{e}$ for valence and conduction band,
respectively. From that we can calculate the corresponding bandwidths\begin{equation}
W=\frac{6\hbar^{2}}{m^{*}a^{2}}\end{equation}
which leads to W$^{v}$=2.2eV and W$^{c}$=7.4eV and is in agreement
with the approximated widths from the band structure calculations
for the hexaborids\cite{key-hexaborids,key-kunes}. In accordance
with the measured band gap for the non-magnetic (J=0) CaB$_{6}$ in
{[}\onlinecite{key-cab6-gap}{]} we assume an intrinsic gap of $\Delta$=0.25eV.
The chemical potential is determined to result in compensated electrons
and holes since the intrinsic band structure is semiconducting.

\subsection{Q-DOS}

\begin{figure}
\includegraphics[%
  width=0.90\columnwidth,
  keepaspectratio]{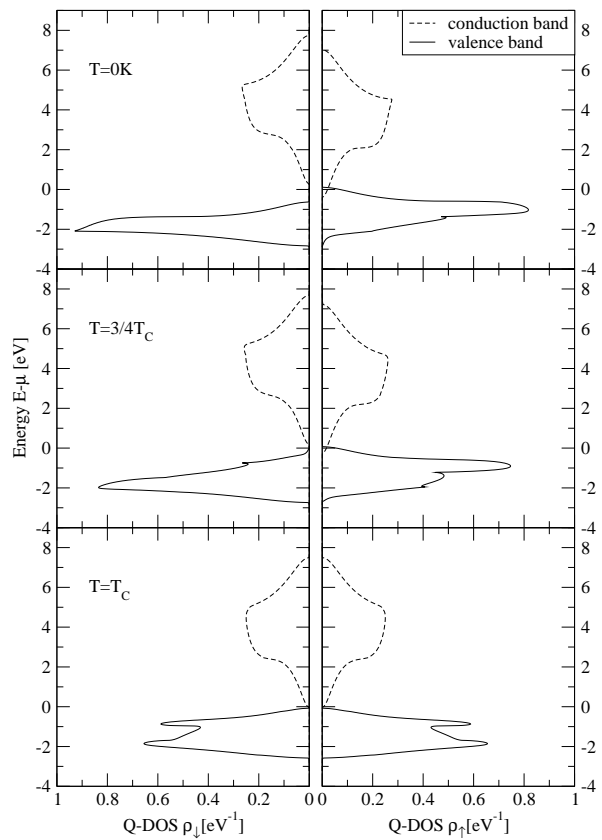}

\caption{\label{cap:DOS}Quasiparticle density of states for spin down (left)
and up (right) electrons in valence (straight line) and conduction
band (dashed line) at T=0K (top), $\frac{3}{4}$T$_{C}$ (middle)
and T$_{C}$ (bottom), S=3.5eV, J=0.2eV, W$^{v}$=2.2eV, W$^{c}$=7.4eV,
$\Delta$=0.25eV.}
\end{figure}
Figure \ref{cap:DOS} shows the quasiparticle density of states (Q-DOS)
for the electrons in the valence and conduction band at different
temperatures (0K, $\frac{3}{4}$T$_{C}$, T$_{C}$) \begin{equation}
\varrho_{\sigma}(E)=-\frac{1}{N\hbar}\sum_{\vec{k}}\frac{1}{\pi}\mathrm{Im}\, G_{\vec{k}\sigma}(E-\mu).\end{equation}
 Since JS$\ll$W we are in the weak coupling regime, although the
effective coupling strength $\frac{JS}{W}$ is different for conduction
and valence band which leads to different results.

For the \textbf{conduction band} the Q-DOS is shifted with magnetization
in a mean-field like manner by $\frac{1}{2}JM_{\sigma}$. Even so
one can see an overall band broadening with decreasing magnetization.

For the \textbf{valence band} some correlation effects are visible.
At T=0K the spin down band is rigidly shifted by $\frac{1}{2}$JS.
Since the local moments are completely aligned no spin exchange is
possible for spin down electrons (keep the anti-ferromagnetic coupling
in mind) hence there is no deformation of that band. In the spin up
channel one sees a dip between two overlapping subbands. The lower
subband describes electrons which flip their spin during the interaction
with the local spin system, emitting a magnon. Thus their lifetime
is decreased. The upper subband is connected to a stable quasiparticle
known as magnetic polaron which propagates through the lattice dressed
by a virtual cloud of magnons (for more details see {[}\onlinecite{key-lowdensity}{]}).

At higher temperatures the scattering for both spin orientations becomes
more likely, explaining the deformation of the Q-DOS in both spin
channels. With decreasing magnetization (increasing temperature) the
difference between the up and down spectrum decreases and vanishes
at T$_{C}$ due to the missing alignment of the local moments.

What is also shown in figure \ref{cap:DOS} is a transition from halfmetallic
to semiconducting behavior undergoing the magnetic phase transition.
The overlap of the spin up bands at 0K is 0.5eV and the gap between
valence and conduction band at T$_{C}$ is 0.04eV. The band gap at
T$_{C}$ is in accordance with tunnel experiments\cite{key-tunnel}.

With a mean-field treatment of the Hamiltonian one could also show
the halfmetal to semiconductor transition. The tight binding dispersion
would be shifted rigidly by $\frac{1}{2}JM_{\sigma}$, resulting in
an overlap at T=0K of 0.45eV for spin up and the band gap at T$_{C}$
of 0.25eV. The smaller band gap within our approach results from the
interaction between the electrons and the local moments even above
T$_{C}$. Although the net magnetization at T$_{C}$ is zero, the
statistically distributed magnetic moments influence the electronic
structure which results in the band broadening. In a meanfield treatment
the interaction is simply switched off at T$_{C}$.

\subsection{Spectral Density}

\begin{figure}
\includegraphics[%
  width=0.80\columnwidth,
  keepaspectratio]{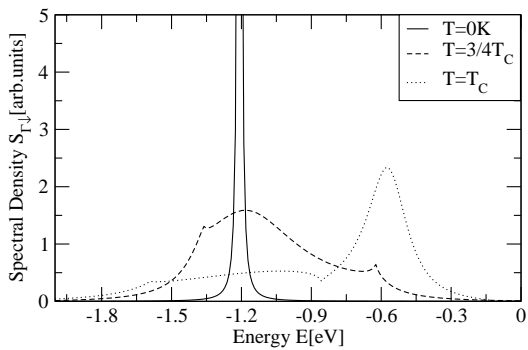}

\caption{\label{cap:spectral-density}Quasiparticle spectral density at the
$\Gamma$-point for spin down quasiparticles in the valence band at
T=0K, $\frac{3}{4}T_{C}$ and T$_{C}$, S=3.5eV, J=0.2eV, W$^{v}$
=2.2eV, W$^{c}$ =7.4eV, $\Delta$=0.25eV}
\end{figure}
The spectral density \begin{equation}
S_{\vec{k}\sigma}(E)=-\frac{1}{\pi}\mathrm{Im}\, G_{\vec{k}\sigma}(E)\end{equation}
 can be compared directly with angle and spin resolved (inverse) photo
emission experiments.

Figure \ref{cap:spectral-density} shows the spectral density at $\Gamma$=(0,0,0)
for a spin down quasiparticle in the valence band at different temperatures.
As explained in the previous section, at zero temperature the spin
down electron in the valence band cannot exchange its spin with the
local moments because they are fully aligned. Thus the quasiparticle
has infinite lifetime. That fact is visualized in figure \ref{cap:spectral-density}
as a sharp peak for T=0K. For higher temperatures and less magnetic
ordering, magnons are excited and can be absorbed by the spin down
electron in the valence band. Therefore the lifetime is decreased
with temperature which is represented by the wide peaks in figure
\ref{cap:spectral-density}.

\begin{figure}
\includegraphics[%
  width=1.0\columnwidth,
  keepaspectratio]{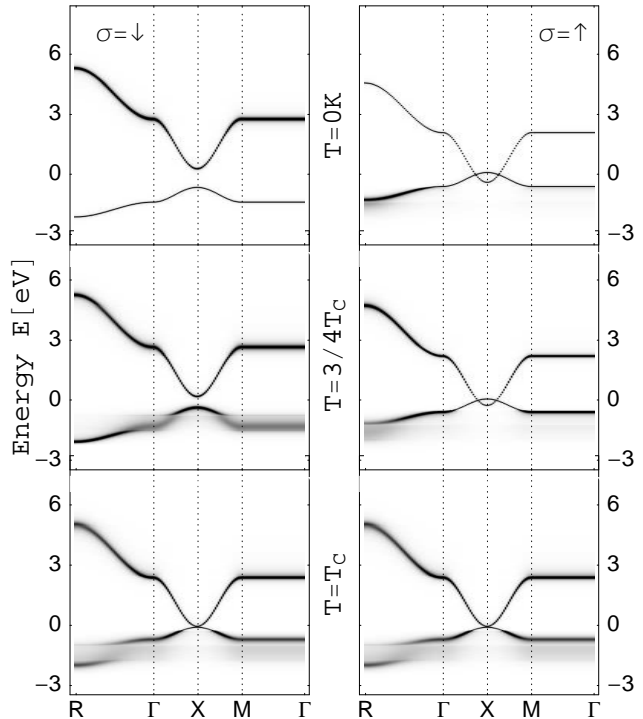}

\caption{\label{cap:band-structure}Spectral density as grayscale at k-points
along the irreducible Brillouin Zone for spin down (left) and up (right)
quasiparticles at 0K (top), $\frac{3}{4}$T$_{C}$ (middle) and T$_{C}$
(bottom), S=3.5eV, J=0.2eV, W$^{v}$ =2.2eV, W$^{c}$ =7.4eV, $\Delta$=0.25eV}
\end{figure}
That evaluation of the spectral density could be done for every k-point.
In figure \ref{cap:band-structure} we show the spectral density as
grayscale along the standard symmetry points in the Brillouin Zone,
which makes a temperature dependent band structure.

The same effects we noticed in the Q-DOS (Fig. \ref{cap:DOS}) can
be seen in the density plots (Fig. \ref{cap:band-structure}). The
spin up conduction band and spin down valence band is undeformed and
rigidly shifted, represented by the thin lines. The smearing occurs
where spin flip scattering is possible.

For example in the plot for spin up quasiparticles in the valence
band at T=0K (fig. \ref{cap:band-structure}, top right, lower band)
we can distinguish between an undeformed, sharp region around the
X-point and a blurred region. These are connected to the two subbands
mentioned in the previous section. The lower one describes quasiparticles
that are involved in spin flip scattering, their lifetime is decreased.
Whereas for quasiparticles in the upper subband no real spin flip
processes are possible because there are no states in the corresponding
spin down region. Their lifetime is infinite.

At higher temperatures the spin flip processes are possible for spin
up and down. It occurs especially in the flat regions of the dispersion
where the possibility of scattering is enhanced due to the smaller
velocity.

Along the $\Gamma-R$ direction occurs a band splitting approaching
T$_{C}$. In a photo emission experiment it would be difficult to
resolve the light shaded regions in figure \ref{cap:spectral-density}
because of the short lifetime of the quasiparticles. Therefore one
might wonder about two bands coming from $\Gamma$ and R which do
not fit together. They can be explained with our many-body theory
of the interaction between the electrons and the local moments.

The effects in the conduction band are less pronounced due to the
weaker effective coupling. Nevertheless one can see that the lines
get smeared with temperature due to the interaction.

\subsection{Plasma Frequency}

Now we will show the qualitative behavior of the effective mass, occupation
number and plasma frequency. That can be compared directly with the
experiments and will be a measure whether our Kondo lattice model
is able to explain some of the properties of EuB$_{6}$.

\begin{figure}
\includegraphics[%
  width=0.90\columnwidth,
  keepaspectratio]{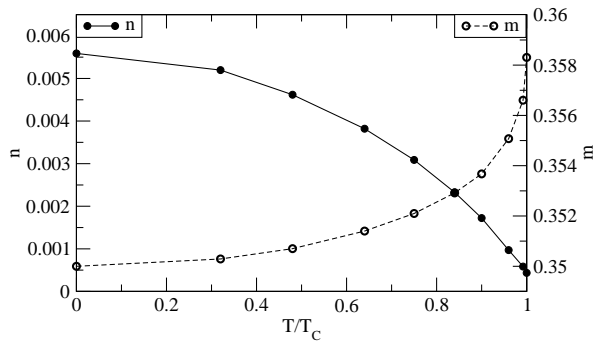}

\caption{\label{cap:mass}Temperature dependence of the effective mass m (dashed
line) and occupation number n (straight line) of spin up quasiparticles
in the conduction band, S=3.5eV, J=0.2eV, W$^{v}$ =2.2eV, W$^{c}$
=7.4eV, $\Delta$=0.25eV.}
\end{figure}

Via the partial derivative of the real part of the self energy R$_{\sigma}$
with respect to the resonance energies $E_{\vec{k}\sigma}=\varepsilon_{\vec{k}}-\mu+R_{\sigma}(E_{\vec{k}\sigma})$
it is possible to derive the effective mass of the quasi-particles\begin{equation}
m_{\vec{k}\sigma}(T)=1-\Bigl(\frac{\partial R_{\sigma}(T,E_{\vec{k}\sigma})}{\partial E_{\vec{k}\sigma}(T)}\Bigr)_{\varepsilon_{\vec{k}}}.\end{equation}
 The effective mass at the X-point for a spin up quasiparticle in
the conduction band is shown in figure \ref{cap:mass} (dashed line).
One can see a slight increase with increasing temperature. That is
in qualitative agreement with the experimental results in {[}\onlinecite{key-degiorgi97}{]}
and {[}\onlinecite{key-paschen}{]}.

The occupation number can be computed by

\begin{eqnarray}
n_{\sigma}(T) & = & \sum_{\vec{k}}\langle a_{\vec{k}\sigma}^{\dagger}a_{\vec{k}\sigma}\rangle(T)\\
 & = & \int f_{-}(T,E)\varrho_{\sigma}(T,E)dE\end{eqnarray}
 where $f_{-}(T,E)=\bigl(1+exp(\frac{E-\mu}{k_{B}T})\bigr)^{-1}$
is the Fermi function (k$_{B}$ is the Boltzmann constant).

Figure \ref{cap:mass} (straight line) shows the occupation number
for spin up quasi-particles versus temperature. In the previous section
the decreasing band overlap with temperature due to the reduced magnetization
was introduced. This is why the occupation number decreases dramatically
by an order of magnitude. This agrees with the Hall data in {[}\onlinecite{key-paschen}{]}.

\begin{figure}
\includegraphics[%
  width=0.80\columnwidth,
  keepaspectratio]{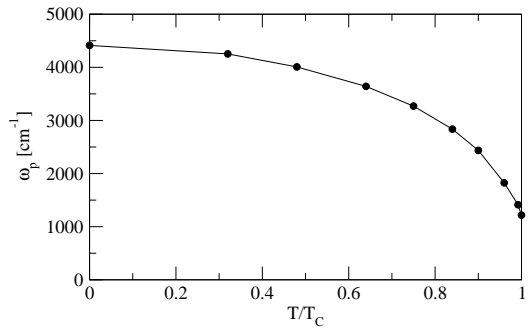}

\caption{\label{cap:Plasma}Temperature dependence of the plasma frequency
for the conduction band.}
\end{figure}

Figure \ref{cap:Plasma} shows the calculated plasma frequency versus
temperature \begin{equation}
\omega_{p}(T)=\sqrt{\frac{e^{2}n(T)}{\epsilon_{0}m(T)}}\end{equation}
with the electric charge e and the dielectric constant $\epsilon_{0}$.
The reduction of free carriers and the increasing effective mass with
increasing temperature lead to a red shift of the plasma frequency.
That is in excellent agreement with magneto optical results in {[}\onlinecite{key-degiorgi97}{]}.

\subsection{Half metalicity}

\begin{figure}
\includegraphics[%
  width=0.80\columnwidth,
  keepaspectratio]{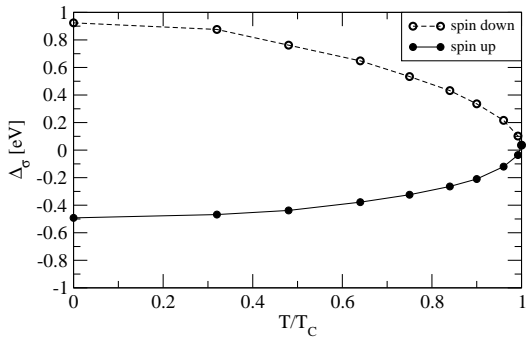}

\caption{\label{cap:Gap}Relative temperature dependence of the band overlap
(straight line) for spin up and the band gap (dashed line) for spin
down quasiparticles.}
\end{figure}

As already mentioned we derive half metallic behavior in the whole
ferromagnetic regime. To clarify that, figure \ref{cap:Gap} shows
the shrinking band overlap for spin up (straight line) and the gap
for spin down quasiparticles (dashed line) versus temperature. At
zero temperature the spin up bands overlap by 0.5eV and the spin down
bands are separated by 0.9eV. With decreasing magnetization the spin
up and down bands are shifted and merge at the Curie temperature,
resulting in a band gap of 0.04eV at T$_{C}$.

\section{Conclusions}

Starting from the two band Kondo lattice Hamiltonian with opposite
sign of the coupling constants for valence and conduction band electrons
we were able to reproduce the metal-insulator transition for EuB$_{6}$
concomitant with the magnetic phase transition. The synchronous band
broadening is verified through our model, too. The experimentally
measured dependence of effective mass, occupation number and plasma
frequency are qualitatively confirmed.

In our model the change in the carrier concentration dominates over 
the change of the effective mass, whereas in the model proposed by Hirsch
\cite{key-hirsch} only the change of the effective mass is made 
responsible for the specific properties of EuB$_6$.

With the chosen parameters, motivated by experiments, 
EuB$_{6}$ turns out to be not only metallic 
but halfmetallic in the ferromagnetic
regime. That would need further experimental investigations since the
halfmetallic behavior is announced as a possible result of future
GW-calculations in {[}\onlinecite{key-kunes}{]} and not explicitly
excluded in the Shubnikov - de Haas and de Haas - van Alphen measurements.

A possible refinement of our calculation will be to consider a real
band structure for EuB$_{6}$. Above T$_{C}$ up to 30K is a regime
of phase separation where bound magnetic polarons and percolation
effects play a significant role. That is not included in our model
yet, since the magnetization is not calculated self consistently.
That should be done in future corresponding to {[}\onlinecite{key-gd}{]}.
Further on it would be interesting to study the diluted system Ca$_{x}$Eu$_{1-x}$B$_{6}$
as was done in {[}\onlinecite{key-pereira}{]}.

\end{document}